\documentstyle[twoside,12pt]{article}
\newfont{\Gigant}{cmr17 scaled 3583}
\newcommand{\slashed}[1]%
{{%
\setbox0=\hbox{$#1$}%
\makebox[0pt][l]{$#1$}%
\makebox[\wd0][c]{/}%
}}
\def \L {{\cal L}}

\newcommand{\lapprox}{%
\mathrel{%
\setbox0=\hbox{$<$}
\raise0.6ex\copy0\kern-\wd0
\lower0.65ex\hbox{$\sim$}
}}
\newcommand{\gapprox}{%
\mathrel{%
\setbox0=\hbox{$>$}
\raise0.6ex\copy0\kern-\wd0
\lower0.65ex\hbox{$\sim$}
}}
\begin{document}
\begin{center}{\Huge Note on Strange Quarks in the Nucleon}
\end{center}
\vskip.5cm
\centerline{K. Steininger\footnote{e-mail:
steininger@physik.uni-regensburg.de} and W. Weise}
\centerline{Institute of Theoretical Physics, University of Regensburg}
\centerline{D-93040 Regensburg, Germany}
\vskip.5cm
\centerline{ABSTRACT}
\begin{center}
\begin{minipage}[t]{12.7cm}
{\small
Scalar matrix elements involving strange
quarks are studied in several models.
Apart from a critical reexamination of results
obtained in the Nambu and Jona-Lasinio model we
study a scenario, motivated
by instanton physics, where spontaneous chiral
symmetry breaking is induced by the flavor-mixing
't Hooft interaction only. We also investigate possible contributions
of virtual kaon loops to the strangeness content of the nucleon
(University of Regensburg preprint TPR-94-04).}
\end{minipage}
\end{center}
\vskip.5cm
Recently much interest has been focused on
the role of strange quarks in
the nucleon.
The observables under discussion are the strange quark spin content of
the proton \cite{jaffe2,ioka}, the strange quark contribution to the
anomalous magnetic moment and the
electromagnetic radius \cite{kama,koepf,cohen}
and the scalar strange quark density of the
nucleon \cite{ioka,kama,haku}.
In this note we will concentrate on the latter one.

A low energy theorem derived from current algebra  relates the
empirical
isospin even
pion-nucleon scattering amplitude at the Cheng-Dashen point to
the $\pi N$ sigma term \cite{wei}.
By analysing  the currently available
experimental data, taking into account the strong
$t$-dependence of the scalar form factor $\sigma(t)$
(with nucleon spinors $u(P)$),
\begin{equation}\label{form}
\sigma(t)\bar u(P^\prime)u(P)
=m_0\langle P^\prime|\bar uu+\bar dd|P\rangle,
\kern.5cm t=(P-P^\prime)^2,
\end{equation}
the authors of ref. \cite{gass1} conclude:
\begin{equation}\label{gass}
\sigma\equiv\sigma(0)=(45\pm8)\kern.1cm{\rm MeV}.
\end{equation}
Here $m_0=(m_u+m_d)/2$ is the isospin averaged current quark mass.
If the nucleon is free of strange quarks,
$\sigma$ should be equal to the matrix element
of the $SU(3)$-octet scalar quark density
\begin{equation}\label{oct}
\sigma_8=m_0\langle P|\bar uu+\bar dd-2\bar ss|P\rangle.
\end{equation}
A straightforward calculation of baryon masses to first order
in the difference of the current quark masses yields
\mbox{$\sigma_8\approx25$ MeV},
while corrections from $SU(3)$-breaking terms of next order
indicate a shift to larger values \cite{gass3},
\mbox{$\sigma_8=35\pm5$ MeV}.
Writing $\sigma=\sigma_8/(1-y)$ one derives
for the ratio $y$ of strange to non-strange pairs in the nucleon:
\begin{equation}\label{sigma2}
y\equiv\frac{2\langle P|\bar ss|P\rangle}
{\langle P|\bar uu+\bar dd|P\rangle}=0.2\pm0.2.
\end{equation}
This result would indicate a significant violation of the Zweig rule,
e.g. of the intuitive assumption that
the nucleon is free of strange quarks, although with a large error.

Several calculations of such strange quark matrix elements
have been carried out in chiral quark models, e.g.
the Nambu \& Jona-Lasinio (NJL) model
\cite{hat,meiss,njl1}.
In these models
the nucleon is composed
of three quasi-particles, the constituent quarks,
which have a non-trivial  structure governed by the
mechanism of spontaneous chiral symmetry breaking.
Since the nucleon consists of constituent $u$- and $d$-quarks,
isospin symmetry implies that one can replace
the nucleon state $|P\rangle$ in (\ref{sigma2}) by a
constituent $u$-quark
state $|U\rangle $, i.e. $y=$\mbox{$2\langle U|\bar ss|U\rangle /$}
\mbox{$\langle U|\bar uu+\bar dd|U\rangle $}.
We will first outline problems in the calculation of scalar
strangeness matrix elements in such models due
to their sensitivity to the
parametrization and the regularization procedure. Next we
study an extreme instanton scenario.
Then we investigate the role of perturbative meson loop
corrections in building up strange quark pairs in the nucleon.

a) \underline{NJL approach.}\\
In the first part we consider the 3-flavor version of the
NJL model \cite{njl1}. It starts from the
effective Lagrangian
\begin{equation}\label{njl1}
{\cal L}_{NJL}=\bar \psi(i\partial^\mu\gamma_\mu-\hat m)\psi
+{\cal L}_4+{\cal L}_6,
\end{equation}
with $\psi=(u,d,s)^t$ and the current quark mass matrix
$\hat m=$diag$(m_u,m_d,m_s)$. The local four-quark interaction
${\cal L}_4$ is symmetric under the chiral $U(3)_L\times U(3)_R$
group:
\begin{eqnarray}\label{njl2}
{\cal L}_4&=&{G_S}\left[(\bar \psi\frac{\lambda_a}{2}\psi)^2
+(\bar \psi i\gamma_5\frac{\lambda_a}{2}\psi)^2\right]+...,\\\nonumber
\end{eqnarray}
where additional terms need not be considered in the present context
since they do not
enter in the expressions for scalar densities at the mean field level.
Here $\lambda_a$ with $a=0,1,...,8$ are the standard $U(3)$ flavor
matrices including
the singlet \mbox{$\lambda_o=\sqrt{2/3}$ diag$(1,1,1)$}.

In nature the axial $U(1)_A$ symmetry is broken
dynamically, presumably
by instantons. A minimal effective interaction,
suggested by 't Hooft \cite{thooft}, which selectively breaks
$U(1)_A$ but leaves the remaining $SU(3)_L\times SU(3)_R\times U(1)_V$
untouched, is a 6-quark interaction in the form of a
flavor-mixing $3\times3$ determinant:
\begin{equation}\label{njl3}
{\cal L}_6={G_D}\left\{{\rm det}[\bar \psi_i(1+\gamma_5)\psi_j]+
{\rm det}[\bar \psi_i(1-\gamma_5)\psi_j]\right\}.
\end{equation}
The effective Lagrangian (\ref{njl1}) with the interaction
${\cal L}_4+{\cal L}_6$
has been used extensively  in the mean field approximation
to study a variety of low energy, non-perturbative phenomena.
A cutoff $\Lambda$ of order 1 GeV is employed
to regularize momentum space (loop) integrals. The physical
picture behind this model is that strong interactions between
quarks operate at low momenta,
i.e. for quark momenta smaller than $\Lambda$, whereas they are
``turned off''
for momenta larger than $\Lambda$.

For sufficiently strong coupling the vacuum undergoes
spontaneous chiral symmetry
breaking (SCSB). Quark condensates $\langle\bar uu\rangle $,
$\langle\bar dd\rangle$ and $\langle\bar ss\rangle$
develop. Current quarks turn into constituent quarks with large
dynamical masses determined by a set of gap
equations. For example,
\begin{equation}\label{gap}
M_u=m_u-G_S\langle\bar uu\rangle -G_D\langle\bar dd\rangle
\langle\bar ss\rangle,
\end{equation}
with the quark condensates ($q=u,d,s$; $N_c=3$):
\begin{equation}\label{cond}
\langle\bar qq\rangle=
-4 N_c i\int^\Lambda
\frac{{\rm d}^4p}{(2\pi)^4}\frac{M_q}{p^2-M_q^2+i\epsilon}.
\end{equation}
The use of
a four-momentum cutoff $\Lambda_4$ gives
\begin{equation}\label{4mom}
\langle\bar qq\rangle=
-\frac{N_c}{4\pi^2}M_q
\left[\Lambda_4^2-M_q^2\ln(1+\frac{\Lambda_4^2}{M_q^2})\right],
\end{equation}
while employing a three-momentum cutoff $\Lambda_3$ yields
\begin{equation}\label{3mom}
\langle\bar qq\rangle=
-\frac{N_c}{2\pi^2}M_q\left[\Lambda_3
\sqrt{\Lambda_3^2+M_q^2}-M_q^2\kern.1cm{\rm arsinh}(\Lambda_3/M_q)\right].
\end{equation}
Using cutoffs in the range $(0.6-0.9)$ GeV and constituent quark masses
$M_q=(0.3-0.4)$ GeV,
the values for the quark condensates
come out typically as
\mbox{$\langle\bar qq\rangle\approx-(250{\rm MeV})^3$}.

SCSB implies the existence of pseudoscalar
Goldstone bosons. In the chiral limit $m_u=m_d=m_s=0$ and with
${\cal L}_6=0$ the whole pseudoscalar nonet of $\pi'$s, $K'$s, $\eta_o$
and $\eta_8$ is massless. In the NJL model these modes emerge as
explicit solutions of the quark-antiquark Bethe-Salpeter equation \cite{njl1}.
Dynamical $U(1)_A$ symmetry breaking by ${\cal L}_6$ of eqn.(\ref{njl3})
gives the singlet $\eta_o$ a non-zero mass. Furthermore, explicit breaking
of chiral $SU(3)_L\times SU(3)_R$ symmetry by bare quark masses $m_s>m_{u,d}>0$
moves all masses of the pseudoscalar nonet to their physical values.
The $\eta-\eta'$ system is reproduced including
its mixing angle $\theta\approx-10^\circ$ (\ref{etawv}).
In the ``standard'' NJL model
the constituent $u$-quark mass remains as a free parameter which
we can choose as  $M_u\approx330$ MeV, about
one third of the nucleon mass.

The constituent quarks are quasi-particles. Their strong interaction
dresses the valence quarks by quark-antiquark polarization clouds,
so that the constituent quarks have non-trivial formfactors.
In particular, the matrix element
of the scalar quark density of flavor $q$
in a  constitutent $u$-quark is given by the Feynman-Hellmann theorem
\cite{feyn},
\begin{equation}\label{fey}
\langle U|\bar qq|U\rangle =\frac{{\partial}M_u}{{\partial}m_q}.
\end{equation}
With eq. (\ref{gap})
we obtain the following explicit expression for the
$\bar ss$ content of the $u$-quark quasi-particle:
\begin{equation}\label{ssinu}
\langle U|\bar ss|U\rangle =-G_S\frac{\partial \langle\bar uu\rangle }{\partial
m_s}-
G_D\left[\langle\bar ss\rangle\frac{\partial\langle\bar dd\rangle}
{\partial m_s}+
\langle\bar dd\rangle\frac{\partial \langle\bar ss\rangle}{\partial m_s}
\right].
\end{equation}
Note that the important term is the last one proportional to
$\partial \langle\bar ss\rangle/\partial m_s$. Its coefficient $G_D<\bar dd>$
is
well constrained by the physics of the $\eta-\eta'$-system. However, the
value of the derivative
\begin{equation}\label{valder}
\frac{\partial \langle\bar ss\rangle}{\partial m_s}=
\frac{\partial\langle\bar ss\rangle}{\partial M_s}\frac{\partial M_s}{\partial
m_s}
\end{equation}
depends strongly on the regularization procedure. Using
expression (\ref{4mom}) with a four-momentum cutoff one finds that
$\langle\bar qq\rangle$ exhibits a minimum at $M_q/\Lambda_4\approx0.75$. Hence
with
typical cutoffs $\Lambda_4\approx 0.8$ GeV and
$M_s\approx(0.5-0.6)$ GeV, the derivative (\ref{valder}) is extremely small,
and the resulting ratio of strange to non-strange pairs turns out to be
$y<0.02$
in this case, with a strong sensitivity to the parameter
$M_s/\Lambda_4$ \cite{meiss}.
In contrast the three-momentum cutoff \cite{hat}
generally leads to much larger
scalar strange pair admixtures than the four-momentum cutoff \cite{njl1}.
In table I we show typical results obtained with a three-momentum cutoff.

\begin{table}
\begin{eqnarray*}
\begin{array}{|c|c|c|c|c|c|}\hline
           & G_S         &  G_D\langle\bar ss\rangle & M_u & M_s & \Lambda \\
           & ({\rm GeV}^{-2}) & ({\rm GeV}^{-2}) & ({\rm GeV}) & ({\rm GeV}) &
({\rm GeV})\\\hline
{\rm NJL}  & 16.4        & 4.1 & 0.33 & 0.52 & 0.65 \\
{\rm INS}  & 0  & 36.3  & 0.54 & 0.64 & 0.57 \\\hline
\end{array}
\end{eqnarray*}

\begin{eqnarray*}
\begin{array}{|c|c|c|c|c|}\hline
           &\langle U|\bar uu|U\rangle  & \langle U|\bar dd|U\rangle  & \langle
U|\bar ss|U\rangle  & y \\\hline
{\rm NJL}  &   2.2        & 0.6           & 0.2           & 0.15  \\
{\rm INS}  &   1.4        & 0.6           & 0.6           & 0.6   \\\hline
\end{array}
\end{eqnarray*}
\underline{\sc Table I:} {\it Upper part:} input for the NJL and the instanton
(INS)
model with three-momentum cutoff. The parameters are
adjusted to reproduce quark condensates and the
pseudoscalar meson spectrum including decay constants.
{\it Lower part:} resulting scalar density matrix elements for the constituent
$u$-quark
and the strange pair fraction
\mbox{$y=2\langle U|\bar ss|U\rangle /\langle U|\bar uu+\bar dd|U\rangle$}.
\end{table}

b) \underline{Instanton approach.}\\
Some years ago Diakonov and Petrov \cite{dipe} suggested that
SCSB in QCD may be
entirely generated by instantons. If so, the 't Hooft
interaction (\ref{njl3}) should dominate in the
low energy regime of QCD.
Motivated by their work we have studied a version of the
NJL model in which $\L_4$ vanishes ($G_S=0$) and
SCSB is generated only by the flavor mixing term
proportional to $G_D$ in (\ref{gap}).
A careful
search for minima of the effective potential\footnote{The
effective potential of the NJL model is worked out in detail in \cite{njl2}
and reads for \mbox{$G_S=0$}:
$\epsilon(M_u,M_d,M_s)=\sum_{q=u,d,s} w_q^0-G_D\langle\bar uu\rangle
\langle\bar dd\rangle\langle\bar ss\rangle$,
with its free part:
$w_q^0=-(M_q-m_q)\langle\bar qq\rangle+
N_ci{\rm tr}\int\frac{{\rm d}^4p}{(2\pi)^4}\ln\left(-i\not\kern-.07cm
p-M_q+i\epsilon\right)$.} in the
three-momentum regularization scheme shows that this particular
pattern of SCSB requires a coupling strength which
comes out quite stable around $G_D\approx140\cdot\Lambda^{-5}$.
After fixing the pion decay
constant to its physical value $f_\pi=93$ MeV by adjusting
the cutoff at $\Lambda\approx0.57$ GeV,
the constituent quark masses turn out to be rather large, namely
$M_u\approx0.5$ GeV and $M_s\approx0.6$ GeV.
Note that the cutoff for the \mbox{'t Hooft} interaction should be related to
the average instanton size $\rho$
(in fact it should be compared with $1/\rho\approx0.6$ GeV from \cite{dipe}).
Current quark masses and quark condensates are in quite good agreement with
standard values once the empirical
pion and kaon masses are reproduced.
The $\eta$-meson mass $m_\eta\approx0.57$ GeV
comes out close to its empirical value. However, in the $\eta^\prime$-channel
the t'Hooft interaction is repulsive and cannot generate a bound state.

We now present numerical results for the NJL and the pure instanton scenario
(INS) in comparison
(see table I). The enhanced flavor mixing of the INS reflects
the much larger strength of
the 't Hooft interaction (\ref{njl3}) as compared to the one in the
``standard''
NJL approach. Hence, if chiral flavor dynamics is dominated by instantons, one
confronts ratios $y$ as large as $1/2$. With more conventional versions of
the NJL model we find ratios $y^<_{\sim}0.15$.
However, the strong dependence on the regularization scheme prohibits
more reliable estimates, so that these numbers should be considered as
upper limits.

c) \underline{Kaon loops.}\\
Our next task is to
consider perturbative corrections to the quark propagator due to the
emission and reabsorption of pseudoscalar mesons.
The admixture of strange pairs in such processes comes from dressing the quark
with
a kaon cloud as shown in figure 1.
We first convince
ourselves that
the shift $\delta M_u$ of the $u$-quark mass from such meson loops
is small so that perturbation theory is justified. Then we use eq. (\ref{fey})
again
and calculate the correction
\begin{equation}\label{corrstr}
\delta\langle U|\bar ss|U\rangle =\frac{\partial(\delta M_u)}{\partial m_s}
\end{equation}
to the strange quark admixture in the constituent $u$-quark from such
mechanisms.
Perturbative corrections to constituent
quark masses have also been investigated
in \cite{shakin} within the framework of a model restricted to $SU(2)$.
Furhermore Koepf et al. \cite{koepf} used
the $SU(3)$-cloudy
bag model for determining
the strange magnetic form factor and the
strange axial charge of the nucleon.
A recent work \cite{zoller}
employs light-cone meson-nucleon vertex functions for
calculating axial form factors of the nucleon following similar
ideas.

\begin{figure}[t]
\vskip4.cm
\underline{\sc Figure.1:} Self-energy of a $u$-quark in the presence of a
kaon cloud.
\end{figure}
In our calculations we use the semi-bosonized version of the
$SU(3)$ NJL model (\ref{njl1},\ref{njl2}) with
$\bar\psi i\gamma_5(\lambda_a/2)\psi$  replaced by the corresponding
pseudoscalar
mesons treated as collective degrees of freedom.
The self-energy of a $u$-quark of  momentum $p$ dressed by a kaon cloud as in
figure 1 is
\begin{equation}\label{loop}\begin{noindent}
\Sigma(p)=i g^2\int\frac{{\rm d}^4q}{(2\pi)^4}
i\gamma_5\frac{\slashed{q}_+ + M_s}{q_+^2-M_s^2+i\epsilon} i\gamma_5
\frac1{q_-^2 - m_K +i\epsilon}\equiv
M_sA(p^2)-\slashed{p} B(p^2),\end{noindent}
\end{equation}
where $q$ is the loop momentum and $q_\pm=q\pm p/2$;
$M_s$ is the mass of the
strange constituent quark and $m_K$ the kaon mass in the intermediate state.
The kaon-quark coupling constant $g$ is given to
leading order in the pseudoscalar meson mass by the Goldberger-Treiman
relation:
\begin{equation}\label{gotr}
g = \frac{M_u+M_s}{2f_K}+{\cal O}(m_K^2),
\end{equation}
where $f_K$ is the kaon decay constant. The latter is related to the quark
condensates and the
current quark masses (in leading order) by the Gell
-Mann, Oakes, Renner (GOR) relation,
\begin{equation}\label{gmor}
(f_Km_K)^2=
-\frac12(m_u+m_s)\left(\langle\bar uu\rangle +
\langle\bar ss\rangle\right)+{\cal O}(m_K^4).
\end{equation}
The scalar functions $A(p^2)$ and $B(p^2)$ are given in terms of loop integrals
\begin{eqnarray}\label{def}
I_1(\mu)&\equiv&2i\int\frac{{\rm d}^4q}{(2\pi)^4}\frac{1}{q^2-\mu^2+i\epsilon}
\end{eqnarray}
and
\begin{eqnarray}
I_2(\mu,\mu^\prime;p^2)
&\equiv&i\int\frac{{\rm d}^4q}{(2\pi)^4}\frac{1}{q_+^2-\mu^2+i\epsilon}\cdot
\frac{1}{q_-^2-{\mu^\prime}^2+i\epsilon},
\end{eqnarray}
as follows:
\begin{eqnarray}\label{dec}
&A(p^2)=-g^2 I_2(M_s,m_K;p^2),&\\\nonumber
&B(p^2)=-\frac12g^2
\left[\left(1+\frac{M_s^2-m_K^2}{p^2}\right)I_2(M_s,m_K;p^2)+
           \frac1{2p^2}\left(I_1(m_K)-I_1(M_s)\right)\right].
\end{eqnarray}
Explicit expressions for $I_{1,2}$ are given in ref. \cite{njl1}.
For consistency we employ the same
regularization procedure
as used in the primary NJL model which generates the constituent quarks
and the pseudoscalar mesons entering in (\ref{loop}).
The perturbative correction from (\ref{loop}) to the constituent $u$-quark mass
becomes:
\begin{equation}\label{mcorr}
\delta M_u= M_s A(M_u^2)-M_u B(M_u^2).
\end{equation}
Numerically one finds $A(M_u^2)\approx0.2$ and $A\approx2B$ so that the
relative
correction
$\delta M_u/M_u$ of the $u$-quark mass
due to the $K^+$ cloud is about 10\%. This justifies the use of perturbation
theory.

In the absence of $U(1)_A$ breaking effects ($G_D=0$ in the gap equation
(\ref{gap})), the
process $u\to(u\bar s)s$ which turns the $u$-quark into a $s$-quark and a
$K^+$-meson
is the only one that contributes to $\langle U|\bar ss|U\rangle $. With
$G_D\ne0$ there are additional non-leading corrections involving pion cloud
contributions
to the quark self-energy as well, through $\bar ss$-dressings of their
constituent quarks,
but they turn out to be negligibly small.

In our estimate of $K^+$ loop effects we therefore use $G_D=0$ and
adjust $G_S$ in (\ref{njl2}), the current quark masses and the cutoff in such a
way
that $ M_u=330$ MeV and the pion mass, the kaon mass and
kaon decay constant $f_K=114$ MeV coincide with their empirical values.
Following  eq. (\ref{fey}) we then calculate
\begin{equation}\label{corr}
\delta\langle U|\bar ss|U\rangle =
\frac{\partial}{\partial m_s}\left[M_s A(M_u^2)\right]
- \frac{\partial}{\partial m_s}\left[M_u B(M_u^2)\right].
\end{equation}
Several effects cooperate in this expression. First the constituent quark mass
$M_s$ grows
with the current quark mass $m_s$ according to the gap equation (\ref{gap}).
Secondly, the $K^+$ mass is related to $m_s$ by eq. (\ref{gmor}) and the
resulting
contribution to (\ref{corr}) is negative. Furthermore, the Goldberger-Treiman
relation
(\ref{gotr}) together with the gap equation implies a positive
derivative of the kaon-quark coupling constant with respect to
$m_s$.
Altogether, the first effect prevails, and we end up with
\begin{equation}\label{res}
\delta\langle U|\bar ss|U\rangle \approx0.03,
\end{equation}
using a three-momentum cutoff scheme. About half of this value results when a
four-momentum cutoff is used.

Diagonal matrix elements such as $\delta$$\langle U|\bar uu|U\rangle $ from
pion and kaon
loops turn ot to be of the same order of magnitude as (\ref{res}). Hence
$\langle U|\bar uu+\bar dd|U\rangle $ is not substantially modified from its
mean field value in table I, calculated by varying the gap equation (\ref{gap})
with respect
to $m_{u,d}$.
Hence the correction from the $K^+$ cloud to the ratio $y$ becomes:
\begin{equation}\label{res2}
\delta y\approx
\left.\frac{2\delta\langle U|\bar ss|U\rangle }{\langle U|\bar uu+\bar
dd|U\rangle }\right.
\lapprox0.03.
\end{equation}

d) \underline{$\eta$ and $\eta'$ loops.}\\
Loops involving the  $\eta$ and $\eta^\prime$ mesons
contribute to $\delta y$ as well, but their effects are small, as
we will now show. First, the contribution of the
$\eta^\prime$ is suppressed due to its large mass.
However, the $\eta$ mass is not much larger than the kaon mass, so
that the corresponding loop correction has to be investigated more
carefully.

The $\eta$ has the following decomposition:
\begin{eqnarray}\label{etawv}
&\eta=\eta_u\left(\bar uu+\bar dd\right)-\eta_s\bar ss,&
\end{eqnarray}
where
$\eta_u=\frac1{\sqrt{3}}\left(\frac{\cos\theta}{\sqrt{2}}-\sin\theta\right)$,
$\eta_s=\frac1{\sqrt{3}}\left(\sqrt{2}\cos\theta+\sin\theta\right)$ in
terms of the $\eta-\eta^\prime$ mixing angle $\theta$. For
$\theta=-10^\circ$ we have $\eta_u\approx0.5$, $\eta_s\approx0.7$ Note
the reduction by a factor $\eta_u$ of the $\eta$ coupling to a $u$-quark as
compared to that of a kaon. Altogether it turns out
that the correction $(\delta M_u)_\eta$ to the constituent
$u$-quark mass due to the $\eta$ cloud is only about 5 MeV (compared to
35 MeV for the kaon cloud). Next, consider
\begin{eqnarray}\label{etas}
&\delta\langle U|\bar ss|U\rangle_\eta=
\frac{\partial}{\partial m_s}(\delta M_u)_\eta=
\frac{\partial (\delta M_u)_\eta}{\partial m_\eta}\frac{\partial
m_\eta}{\partial m_s}+
\frac{\partial (\delta M_u)_\eta}{\partial \eta_u}\frac{\partial
\eta_u}{\partial m_s}
=\frac{\partial (\delta M_u)_\eta}{\partial m_\eta}\frac{\partial
m_\eta}{\partial m_s}+
\frac{2\cdot(\delta M_u)_\eta}{\eta_u}\frac{\partial \eta_u}{\partial m_s}.&
\end{eqnarray}
The first term on the r.h.s. involves the positve derivative of the
$\eta$ mass $m_\eta$ with respect to the strange quark mass together with
the (negative) change of the $\eta$ loop integral when changing $m_\eta$.
This product has a small numerical value (about $-0.01$). The last term
(in which the relation $(\delta M_u)_\eta\propto\eta_u^2$ has been used)
reflects the
dependence of the $\eta-\eta^\prime$ mixing pattern on the strange quark mass.
For
$m_s=m_u=0$ we have $\theta=0$. Assuming a linear dependence $\theta\propto
m_s$ we
estimate at $\theta\approx-10^\circ$ with $m_s\approx 130$ MeV:
\begin{eqnarray}\label{etass}
\frac{2(\delta M_u)_\eta}{\eta_u}\frac{\partial \eta_u}{\partial m_s}\approx
-10{\kern.1cm\rm MeV}\cdot
\frac{\sin\theta+\sqrt{2}\cos\theta}{\cos\theta-\sqrt{2}\sin\theta}\cdot
                    \frac{\partial\theta}{\partial m_s}
\approx0.013.
\end{eqnarray}
Adding up both terms in eq. (\ref{etas}) gives a negligibly small contribution
to
$\langle U|\bar ss|U\rangle$, about one order of magnitude smaller than that
from the kaon cloud.

In summary we have analysed strange quark admixtures to the scalar density of
the nucleon in terms
of $\bar ss$-components in the quasi-particle structure of
constituent quarks. In our approach such components arise from the
non-perturbative
dressing of the quarks by scalar mean fields, and from perturbative kaon cloud
effects.
We find that in the
``standard'' NJL model with moderate axial $U(1)$ breaking, the upper limit for
the
ratio $y$ of $\bar ss$ pair admixtures in the nucleon relative to $\bar uu$ and
$\bar dd$ pairs is
about 0.15 (with substantial uncertainties due to the strong dependence on
details of
the regularization procedure). In contrast, if instantons dominate the low
energy dynamics
so that the effective interaction is governed by axial $U(1)$ breaking, the
resulting
$\bar ss$ admixtures can be much larger, but with similar uncertainties. Kaon
cloud effects
alone, on the other hand, would not give large strange pair components
in the nucleon.
We find an upper limit of about 3\% from this source.

\end{document}